\documentstyle[amsmath,aps,multicol,prl,epsfig]{revtex}
\newcommand{\ve}{\varepsilon}
\newcommand{\nit}{\noindent}

\begin{document}
\begin{title}
{\bf Breathers on a Background: Periodic and Quasiperiodic
  Solutions of Extended Discrete Nonlinear Wave Systems }
\end{title}

\author{P.G. Kevrekidis\footnote{To whom correspondence should be addressed: \\
Phone: 413-577-1977, fax: 413-545-1801, e-mail: kevrekid@math.umass.edu.}
}
\address{Department of Mathematics and Statistics, University of 
Massachusetts,
Lederle Graduate Research Tower, Amherst, MA 01003-4515, USA} 
\author{M.I. Weinstein}
\address{Mathematical Sciences Research, Bell Laboratories, Lucent 
Technologies,  
 600 Mountain Ave., Murray Hill, NJ 07974-0636}
\date{\today}
\maketitle

\begin{abstract}
In this paper we investigate the emergence of time-periodic and and 
time-quasiperiodic
 (sometimes infinitely long lived and sometimes very long lived 
or metastable)
  solutions of discrete nonlinear wave equations: discrete sine Gordon, discrete
 $\phi^4$ and discrete nonlinear Schr\"odinger. The solutions we consider
are periodic oscillations on a kink or standing wave breather background.
The origin of these oscillations is the presence of internal modes,
associated with the static ground state. 
 Some of these modes are associated with the breaking of 
translational invariance, in 
going from a spatially continuous  to a spatially discrete system. Others
are associated with discrete modes which bifurcate 
from the continuous spectrum.
It is also possible that such modes exist  in the continuum limit
 and persist in the discrete case. The regimes 
of existence, stability and metastability of states as the lattice
spacing is varied are investigated analytically and numerically.
A consequence of our analysis is a class of spatially localized, 
time quasiperiodic solutions of the discrete nonlinear Schr\"odinger
equation. We demonstrate, however, that this class of quasiperiodic
solution is rather special and that its natural generalizations yield
only metastable quasiperiodic solutions.
\end{abstract}


\begin{multicols}{2}

\section{Introduction}

In the past two decades, there has been an explosion of interest
in the study of solitary waves in spatially discrete nonlinear systems.
Prominent examples, where one expects important phenomena to be 
intrinsically associated with spatial discreteness, are: the
motion of dislocations in solid-state physics \cite{FK,AC}, the
transmission of kink-like bits of information in Josephson junction
arrays \cite{UDVPHO}, the propagation of pulses in
nonlinear optical waveguides \cite{KHKLK1,KHKLK2} (where they
present great potential for applications such as all-optical
switching, steering and demultiplexing  see e.g., \cite{DKL} and references
therein), or even
in chemical \cite{SBLSSBZS} and biological \cite{PB} (and references
therein) applications. 
 Recent experimental work
 confirming the relevance of such models has  appeared in
 \cite{SBLSSBZS,CaGia,dnlsexperiment}.

The models arising in many of these studies, 
governing the behavior of a field $u_n(t)$ depending 
on a continuous time variable $t$ and discrete spatial variable $n$, are: 

\nit{\it the discrete SG equation} (DSG)
\begin{eqnarray}
\ddot{u}_{n}=u_{n+1}+u_{n-1}-2 u_n- \frac{1}{d^2} \sin{u_n},
\label{tprleq1}
\end{eqnarray}

\nit {\it the discrete $\phi^4$ equation} ($D\phi^4$)
\begin{eqnarray}
\ddot{u}_{n}=u_{n+1}+u_{n-1}-2 u_n +\frac{1}{d^2} (u_n-u_n^3),
\label{tprleq2}
\end{eqnarray}

\nit and the {\it discrete nonlinear Schr\"odinger equation} (DNLS)
\begin{eqnarray}
i \dot{u}_{n}=-C (u_{n+1}+u_{n-1}-2 u_n)- |u_n|^2 u_n.
\label{tprleq3}
\end{eqnarray}
We have used the standard notation where $d=1/\Delta x$ is the inverse spatial 
 lattice
spacing and  $C=d^2=(\Delta x)^{-2}$ is the coupling constant.

An important phenomenon associated with spatial discreteness is the role of
the spatial lattice sites  as ``defects'' to which propagating states can
be ``trapped'' or ``pinned'' \cite{PK,CY,KEVWEI}. 
At a structural level, this is a result of the breaking of
translational invariance.
Indeed, the spatially continuous variants are Lorentz invariant (SG, $\phi^4$)
 and Galilean invariant (NLS); a stationary coherent structure can be 
``boosted", i.e., transformed, into one traveling with a constant velocity.

The main coherent structures of interest in these equations
(see {\it i.e.},
  \cite{KEVWEI,BCK,KJ,JAUB,KRB,KRB1}) are kink-like excitations in DSG
and $D\phi^4$  and time periodic, exponentially localized in space pulse like
excitations in DNLS (solitary waves). The latter
  are frequently referred to as {\it discrete breathers}.

In the analytical and numerical work on these discrete systems 
\cite{PK,CY,JAUB,BCK} 
 two main steady states have been identified;
 there are actually many other steady states \cite{BCK,AUB}.
 One is centered on a lattice site and one is centered between
 consecutive lattice sites. In the case of DSG and $D\phi^4$ kinks,
the stable configurations (energy minimizers) 
 are those centered between sites, while
for the breathers of DNLS, stable configurations (constrained energy minimizers;
the constraints are imposed by the additional conserved quantities such as
the $l^2$ norm of the solution \cite{THRESH})
 are those pulses which are centered at a lattice site.


Until recently,
it was believed that an initial condition, which for the 
 translation invariant systems leads to a coherent structure propagating
to infinity at a uniform speed, would in these discrete systems
result in a state radiating energy which would eventually get  trapped
 by the ``Peirels -Nabarro'' potential \cite{PK} associated with  
the lattice  and subsequently approach a time independent state. 
However, it was
established in \cite{KEVWEI} that, depending on the parameter regime
 (lattice spacing), the asymptotic state may, in fact, be time-periodic. It
may also consist of a  time quasiperiodic state which slowly decays 
to a time-periodic state.
 The rate at which
the system relaxes to its asymptotic state is controlled by 
resonances of ``localized internal modes'' with radiation modes.
Internal modes are spatially localized modes associated with the linearized 
operator about the coherent structure with corresponding discrete 
eigenfrequencies (point spectrum), which are not related to an
underlying group invariance (zero modes). 
  They give rise to time 
periodic solutions of the linearized evolution equation, of the
form $\phi_n^{\omega} \exp(i \omega t)$. Radiation modes are those
associated with the continuous spectrum of the linearization.

In a recent paper \cite{KEVWEI}, using a methodology developed in 
 \cite{SW} (see also \cite{PEL1}), we gave a systematic
approach for deriving details of the decay to an asymptotic state.  
We now briefly describe these results. 
 Suppose $\omega_1$ and $\omega_2$ are two internal
mode frequencies associated with a  coherent structure of (\ref{tprleq1}),
(\ref{tprleq2}) or (\ref{tprleq3}). Consider initial conditions which are a 
 small perturbation of the coherent structure and that the perturbation 
(of $O(\ve)$) 
consists of components in the direction of the internal modes, 
 $\phi^{\omega_1}$ and $\phi^{\omega_2}$.  As time evolves, 
 nonlinearity generates various linear 
combinations of these frequencies so we investigate where such combinations
land relative to the so-called phonon band (PB).
Suppose that 
 a combination of frequencies $k\omega_1+l\omega_2$
 falls inside the PB with  $|k|+|l|=n$. 
For DSG and $D\phi^4$, PB is, respectively, the purely 
 imaginary interval $\pm i[1/d,\sqrt{4+1/d^2}],
\pm i[\sqrt{2}/d,\sqrt{4+2/d^2}]$ and  for DNLS it is 
 $\pm i[\Lambda,\Lambda+4 C]$, where
$\Lambda$ is the time-frequency of the ground state discrete solitary
wave.
 In this case we showed in \cite{KEVWEI}
 that the time evolution of the internal mode 
 powers (squared amplitudes), $P=P_{\omega_1}$ and  $Q=Q_{\omega_2}$ 
 is controlled by a system whose
normal form is
\begin{eqnarray}
P_t = - \sum_{k,l \in I^{res}} \ve^{2(|k|+|l|-1)}  \Gamma_{1;k l} P^k Q^l
\label{tprleq4}
\\
Q_t = - \sum_{k,l \in I^{res}}  \ve^{2(|k|+|l|-1)}  \Gamma_{2;k l} P^k Q^l,
\label{tprleq5}
\end{eqnarray}
where $I^{res}=\{k,l: k \omega_1+l \omega_2 \in PB\}$ is the set of
resonant frequency combinations. The 
 ``damping coefficients''   
$\Gamma_{1; k l}, \Gamma_{2; k l}$ can be computed \cite{KEVWEI}, are
always nonnegative  
and are generically strictly positive.
 From (\ref{tprleq4})-(\ref{tprleq5}), one concludes that the
decay rate is controlled by the appropriate harmonic e.g., if 
$ 2 \omega_1 \in $ PB, then the (long time asymptotic) 
decay of $P$ goes as $t^{-1/2}$, if
$3 \omega_1 \in$ PB, it goes as $t^{-1/4}$ and so on.
  As the parameters
of the problem (such as the lattice spacing) vary, the set
of resonances $I^{res}$ may vary
and therefore so will the normal form equations and the nature of the
asymptotic state dynamics; see \cite{KEVWEI} for examples of such 
transitions.
  
\section{Asymptotic states: stationary and periodic, stable and metastable}

What are the possible asymptotic states ? Consider the discrete nonlinear
 wave equations  (\ref{tprleq1})
and (\ref{tprleq2}). There is, of course, the ground state kink but one
may ask whether there are additional time-periodic or even 
quasi-periodic states associated with the time-periodic internal modes.,
e.g., solutions of the nonlinear equation of the form: $K_{gs} + \sum_{\omega \in
\Omega_{internal}} 
\phi_n^{\omega} \exp(i \omega t) +$ C.C. plus higher order corrections.
 Here,  $K_{gs}$ denotes
  the ground state or least energy
kink and $\Omega_{internal}$ denotes the set of localized
internal mode frequencies. In \cite{KEVWEI} we prove the existence of
time-periodic solutions (only {\it one} contribution to the above sum is
present) in certain parameter regimes. No quasi-periodic solutions 
exist since for any internal mode frequencies $\omega_1,\omega_2$,
there always exist integers $k_1 \neq 0, k_2 \neq 0$ such that $k_1 \omega_1
+ k_2 \omega_2 \in$ PB. There are however metastable, 
  very long-lived
states of this type.

For DSG, depending on parameter regime, we have between one 
and three possible asymptotic states from among $K_{gs}$ (the ground
state kink), $gW$ and $eW$, each with their own basin of attraction.
$gW$ and $eW$ are time-periodic solutions associated with perturbed 
translational modes
(i.e., bifurcating from the origin of the spectral plane) 
and edge (bifurcating from the edge of the continuous spectrum) 
 internal modes (Theorem 5.1 of \cite{KEVWEI}). See Table I.

For $D\phi^4$, we have between one and three from among 
$K_{gs},\  gW,\  W$ and $eW$. The additional periodic solution $W$, is
associated with the additional {\it shape mode} present in the point spectrum
of the $\phi^4$ model. See Table II.

The above periodic solutions are localized time-periodic excitations on a kink
(spatially inhomogeneous) background, ``breathers on kinks'' (BOK's); see Fig. 
1. Their dynamic stability in various regimes of the discreteness 
parameter $d$ was numerically investigated in \cite{KEVWEI}.

We now turn to DNLS. Seek solutions of the form
\begin{eqnarray}
u_n(t)=\exp(i \Lambda t) \psi_n.
\label{Ansatz}
\end{eqnarray}
This gives rise to a system of nonlinear algebraic equations
for $\{\psi_n\}$
\begin{eqnarray}
-\Lambda \psi_n=-C (\psi_{n+1}+\psi_{n-1}-2 \psi_n) -|\psi_n|^2 \psi_n.
\label{BNLS}
\end{eqnarray}
A variational approach to the questions of existence, excitation thresholds
and nonlinear Lyapunov stability of the
 ground states of (\ref{BNLS}) is presented in  \cite{THRESH}.
Solutions can also be constructed by iterative methods 
implemented numerically in \cite{KRB,KRB1}.
 Linear stability analysis proceeds by
seeking a solution of the form 
\begin{eqnarray}
u_n(t)= \exp(i \Lambda t) (\psi_n + v_n(t))
\label{tprleq6a}
\end{eqnarray}
where $v_n=a_n \exp(- i \omega t)+ b_n \exp(i \omega t)$ denotes
the perturbation about the ground state. This leads to the spectral problem

\begin{eqnarray}
\omega a_n=-C \Delta_2 a_{n}- 2 |\psi_{n}|^2 a_{n} +
\Lambda a_{n}-{\psi_{n}}^2 {b^\star_{n}},
\label{tprleq7}
\\
-\omega b_{n}=-C \Delta_2 b_{n}- 2 |\psi_{n}|^2 b_{n} +
\Lambda b_{n}-{\psi_{n}}^2 {a^\star_{n}}.
\label{tprleq8}
\end{eqnarray}
The analysis of the modes for different values of the relevant parameter 
$\Lambda/C$ is given in \cite{JAUB}.
There it is found for the stable 
(site-centered) pulses of
DNLS that there are two modes, a perturbed translational internal mode (the
``pinning'' mode) and an edge mode (the ``breathing''
mode). The picture for DNLS is derived in \cite{JAUB} and is analogous to
that described above and in \cite{KEVWEI} 
 for equations (\ref{tprleq1})-(\ref{tprleq2}). 
In fig. 2, the
relevant (localized) eigenmode frequencies are given and compared
with those of the PB. The analysis of 
\cite{JAUB} shows that the methodology of \cite{KEVWEI} can be
applied in this case, even though the basis of eigenfunctions
is not orthonormal  (the breathing mode has non-vanishing overlap
with the rotational symmetry
eigenmodes with $\omega=0$). However, a pseudo-scalar product
can be used \cite{JAUB} in the spirit of \cite{Kaup} to perform
the same analysis. Johansson and Aubry do this in two alternative
ways (higher order perturbation theory and a conservation law approach)
but nonetheless reach a special case of the result of \cite{KEVWEI}
proving that when the $p$-th harmonic of a mode is inside PB, then
the amplitude of the mode decays as $|a(t)| \sim t^{-1/(2p-2)}$ 
asymptotically, as is suggested also by eqns. (\ref{tprleq4})-(\ref{tprleq5}).
However, in the DNLS case, they found, additionally, that 
the existence of the additional two-dimensional
manifold of eigenmodes with zero frequency, in order to avoid unphysical
divergences of higher order perturbations (due to projections on this
manifold), causes an increase of
the breather mode frequency according to $\Lambda_t \sim |a|^{2p}$.

The methods of \cite{KEVWEI} can be used to extend the picture developed
 in \cite{JAUB}. Using the information, displayed in figure 2, 
  on the location
of linear combinations of internal mode frequencies associated with the
ground state of equation (\ref{BNLS}), relative to the phonon band PB,
we can conclude: 
\begin{itemize}
\item For any value of $\Lambda/C$ for which the edge mode
exists ($0<\Lambda/C<1.7$), initial conditions
close to it decay as $t^{-1/2}$ since
its second harmonic is always inside PB. 
\item 
The perturbed translational mode (which disappears around $\Lambda/C\sim 1.1$
as is noted in \cite{JAUB}), for $0.48<\Lambda/C<1.1$, $2 \omega_g$ is
inside PB, hence the decay in this range goes
  as $t^{-1/2}$. However, for $0.3997<\Lambda/C<
0.48$, the third harmonic resonates with PB and hence the 
decay goes as  $t^{-1/4}$.
Subsequently $t^{-1/6}$ for $\Lambda/C \in[0.3576,0.3997]$ and 
$t^{-1/8}$ for $\Lambda/C \in[0.3304,0.3576]$ are implied by 
 principle resonance occurring for higher   
harmonics.
\end{itemize}

Therefore, for any value of the coupling there are 
harmonics of the individual modes that are in the phonon band and hence
cause decay of the internal mode oscillations. 
 Hence, no genuinely periodic solutions along eigendirections
can survive in this case (in the frame co-rotating with the discrete
breather frequency $\Lambda$). The decay however can be extremely slow as we 
now explain. The perturbed 
translational mode, the neutral oscillatory mode to which the translation 
zero mode  is perturbed for $\Delta x>0$, 
 approaches zero very fast for small $\Lambda/C\sim \Delta x$, 
 the continuum
limit. This approach has been shown
  to be exponentially fast as $\Delta x \downarrow 0$ 
($\omega^2 \sim \exp(-\pi^2/\Delta x)$) in \cite{KKJ,KK}. 
Therefore, as seen in figure 2 and can be derived from the above
estimates, for very small values of $\Lambda/C$, only {\it very}
high harmonics will fall inside PB.  
 By the  analysis of 
  \cite{KEVWEI} this corresponds to very slow time decay rates of the 
internal mode oscillations, which may be hard to detect numerically.
 An example of this sort is given in figure 3. 
The numerical experiment is performed for DNLS, with $\Lambda/C=0.370$
and for the  perturbed translational
mode
  of $\omega/C=0.102$, the $4$-th harmonic 
is the leading one in the phonon band. The initial condition is taken to be
 the exact breather profile with a small perturbation along the perturbed
 translational mode  eigendirection.
  The 
 background oscillations are clearly discernible
 as time-dependent side wobbles in the evolving profile.
Our simulations show very slow decay of these oscillations, in 
accordance with the theoretical prediction. 
The analysis of  \cite{KEVWEI} implies that 
  these modes will decay to the ground state standing wave breather
but only over very long time scales.
Modes such as the one that is observed in figure 3 are very long-lived 
time-periodic and spatially localized.  
In view of (\ref{tprleq6a}) they correspond to slowly decaying (metastable)
 quasi-periodic oscillations  of DNLS (which are 
periodic in the rotating frame). The periodic oscillation is mounted on a 
breather background.
 Hence these modes are (metastable) Breathers On Breathers (BOB's).
We note  that there are other constructions 
of quasiperiodic modes  in discrete systems. 
Such attempts (see, for example,
\cite{JAUB2}) exploited continuation from the anti-continuum limit of
two oscillators with different frequencies. Also, for polaron
type models such as the Holstein model, a proof of existence of
quasiperiodic solutions was give in \cite{AUB}.
Our analysis above is not restricted 
to the regime where the sites are weakly coupled. However, it
should be noted that very close to the continuum limit, logarithmic
effects similar to the ones described in \cite{PEL2} for generalized
{\it continuum} NLS equations may become relevant. The work presented
herein mainly focuses in the regimes of large or moderate
discreteness and the mechanisms described will still be present
even for very small discreteness. The work of \cite{PEL2}, however,
raises the interesting question as to whether there is a crossover
point beyond which (in approaching the continuum limit) logarithmic
effects become dominant over power law ones. Such issues will be
addressed in a future publication.
\medskip

\section{\bf Infinitely long-lived quasiperiodic solutions}
Are there time quasi-periodic states which are non-decaying?
We now display a family of (non-decaying) time quasi-periodic and spatially
localized BOB states. Our point of departure will be the modes introduced in 
\cite{DKL}, which we call twisted localized modes (TLM's). 
These modes are solutions of DNLS
of the form (\ref{Ansatz}) with $\{\psi_n\}=(\dots,\alpha,1,-1,-\alpha,\dots)$
with $0<\alpha<<1$. They can be viewed as soliton-antisoliton bound
pairs which can exist due to discreteness \cite{KKM,Kprep}. There is no
continuum analog since in the continuum case a soliton-antisoliton
initial condition naturally repels.

Henceforth, for simplicity
 we will fix $\Lambda=1$ and vary $C$ only.
We  construct  these modes numerically by solving equation 
 (\ref{BNLS})
and subsequently solve numerically eqs. (\ref{tprleq7})-(\ref{tprleq8})
to determine 
their linear stability. We find that
\begin{itemize}
\item TLM's are stable for $1/C \leq 0.146$. Their spectrum for 
this parameter range consists of two modes at $\omega=0$ and an additional
internal mode with frequency $\omega_{\star}$ in the gap between $0$ and edge of PB.
\item For values $1/C>0.146$ a series of instabilities arise giving
quite interesting phenomena that have been analyzed elsewhere \cite{Kprep}.
\end{itemize}
For $C$ in the range of stability, the location of multiples of the single
non-zero frequency ($\omega_{\star}$) 
relative to  PB
is displayed in figure 4. It can be seen that 
\begin{itemize}
\item For $1/C \in [59.712,67.681]$, $4 \omega_m \in$ PB and hence
implying decay of the internal mode oscillation about the TLM 
 with a rate:
  $t^{-1/6}$. For $1/C \in [31.943,39.604]$, $3\omega_m \in$
PB implying $t^{-1/4}$ decay and for $1/C \in [11.989,19.288]$, 
 $2\omega_m \in$ PB
 implying a $t^{-1/2}$ decay rate. 
\item More interestingly, as figure 4 suggests, the gaps between
 the above coupling parameter intervals
 constitute parameter regimes  for
which no multiples of $\omega_{\star}$ lie inside PB. Hence 
an initial condition consisting of the exact mode plus a perturbation 
along the $\omega_{\star}$ eigendirection encounters no evident decay 
mechanism.
Indeed, the proof of Theorem 5.1 of \cite{KEVWEI} can be carried through
in this case
to obtain the existence of time-periodic solutions of the nonlinear
perturbation equation governing $v_n(t)$. In this way, we have 
constructed genuine time quasi-periodic solutions (BOB's) of DNLS
of the form (\ref{tprleq6a}), where $v_n(t)$ is time periodic.
 This is  demonstrated
in figure 5, where we display the results of a  numerical experiment for 
 $C=0.02,\Lambda=1$, where the internal mode frequency is 
 $\omega_{\star}=0.294$. We observe
the very regular oscillations on top of the breather amplitude due to
the BOB's quasi-periodicity. 
\end{itemize}

\section{Generalization to Multipulses}

We saw in the previous section that TLM's can be thought of as
a pulse and an antipulse whose ``backs'' have merged. The question
that then stems from this analogy is whether genuinely quasiperiodic
solutions of DNLS arise naturally in such multipulse setups where
two or more pulses have been concatenated.

A general analysis of such configurations for discrete setups has
been given in \cite{KKM} for two pulses and generalized in \cite{Knew}
for multiple pulses. In these works it has been found that in order
for multipulses to be stable, the nearest neighbor ones have to
possess opposite parity, i.e., a phase difference of $\pi$.
Hence, we will consider only such configurations hereafter.

In order to understand the possibilities for quasiperiodic 
solutions of multipulse type, let us return to the linearization
picture for a single pulse. As is well known, for the single pulse,
there are two pairs of eigenvalues near the origin of the spectral
plane, the rotational modes (of zero frequency) and the translational
modes of frequency dependence (on $h$) as $\exp(-\pi^2/h)$. For a
concatenation of N pulses, there will be $2 N$ pairs near the
origin. $N$ of them will be translational (stemming from the
linear combinations of the translational eigenmodes of the 
individual pulses), 1 pair will be the rotational one of $\omega=0$
and $N-1$ pairs will pertain to the so-called interaction eigenmodes
\cite{KKM,Knew}, whose frequencies will, in general, behave
as $\exp(-L)$, where $L$ is the (typical) pulse separation.
It should be noted that for such multipulse configurations to
occur, it necessary for the pulses to have a separation larger
than a minimal one $L>L_{min}$.

For values of $h \leq O(1)$, as was found in \cite{KKM,Knew},
it is generically true that for $L>L_{min}$, the interaction
eigenvalues have frequencies typically lower than the translational ones.
Hence, since as we saw in section II, quasiperiodic solutions cannot
be formed for such regimes of the lattice spacing, as the
translational eigenvalues resonate with the continuous spectrum,
hence such solutions will also not be but metastable for the
multipulse configurations as well.

From the above argument, it can be seen that for multipulse 
configurations, as was the case with the TLM's of section III,
only strong discreteness contexts $h >> 1$ provide the possibility
for quasiperiodic solutions. In such regimes of strong discreteness,
the translational eigenmodes have already merged with the continuous
spectrum \cite{JAUB} and only the interaction eigenmodes (and their
harmonics) relative position to the continuous spectrum has to be
accounted for.

In order to understand the behavior of multipulses, a natural
continuation parameter that we will use in this context is the
separation between the pulses $L$. In the case of the TLM's,
this separation will be considered $h$, i.e., the distance
between the pulse maximum and the antipulse minimum. In this
case, the form of the solution is shown in the top left panel
of Fig. 6 and the form of the evolution of $\omega/C$ as a function
of $\Lambda/C$ is given in the top right panel. In the middle
two and lower two panels of fig. 6, similar results are shown
for $L=2 h$, and $L=3 h$ respectively. What can be clearly
observed is that in these cases, the dependence of the internal
(interaction) eigenmode on the coupling constant is such that
{\it for any value of $h$}, there are harmonics of the internal
mode frequency {\it inside} the PB. Hence, the TLM's are very
special in the sense that the relation of their centers positions
is such that it permits for a set of intervals of coupling constants
to have no resonances of internal modes with the PB and, thus,
quasiperiodic solutions.

To show the generality of the above conclusion, we performed
similar numerical experiments for $3$-pulse configurations
(in which the middle one was of opposite phase than the two
outer ones). The case of $L=h$ is once again shown in the
top left panel of Fig. 7 and the corresponding resonance diagram
of $\omega/C$ as a function of $\Lambda/C$ appears in the top right
panel of the same figure. The same plots for $L=2 h$ and $L=3 h$
are shown in the middle and lower rows of Fig. 7 respectively.
In this case, as per the general discussion presented above,
there are two pairs of interaction eigenmodes. In the case of
$L=h$, it can be observed that their frequency increases as 
a function of $\Lambda/C$ and hence gaps appear in the presence
of the modes' harmonics in the PB. Such gaps for instance are
found for $h \in (4.677,5.292)$, $h \in (6.537,7.482)$ for the
first frequency (in solid line) or for 
$h \in (4.321,6.468)$ for the second internal mode (in dashed line).
Excitation of one of these internal modes in the respective intervals
will give rise to genuinely quasiperiodic solutions. On the other
hand, it can be observed that already for $L=2 h$, once again the
dependence of $\omega/C$ on $\Lambda/C$ is monotinically decreasing
and that harmonics will always be present inside. The same picture
of course in maintained for $L=3 h$ and for larger interpulse
distances.

These results lead us to the conclusion that generically quasiperiodic
solutions of the TLM variety (and its multipulse generalization) will
be present in DNLS type systems. However, this is a rather
``delicate'' result of the interplay between the lattice spacing $h$
and the interpulse separation $L$ which has been found to be quite
special to such. More general multipulse solutions of larger
interpulse distance are invariably found to possess internal modes
whose harmonics are resonant with the PB and hence in the latter
case only metastable quasiperiodic configurations will be present.

\section{Conclusions}
In this paper, we have given theoretical justification as well as
numerical evidence of the existence and asymptotic stability of
breather-like  modes which consist of (depending on parameter regime)
 infinitely
  long lived or very long lived spatially localized and 
  time-periodic oscillations on a kink background
 (BOK's) or
on a breather background (BOB's). These modes are different from
previously reported ones in that they reside on a non-uniform
(coherent structure)
background rather than on a uniform background. Because of their
relevance to asymptotic states for a wide range of initial conditions
(possibly in experimental situations of interest) further study and 
understanding of these modes as well as analytical
investigation of their stability are challenging questions
worth pursuing. The existence of genuinely quasi-periodic solutions is 
apparently rare for nonlinear conservative wave equations on infinite
spatial domains.
This is in contrast to the case of nonlinear wave equations on finite
domains. The discrete case corresponds to finite dimensional KAM theory.
The continuum case (that of vibrations, strings and membranes) is
studied in \cite{CW,Kuk,Bgn}, where such solutions have
been constructed. The non-generic nature of such quasi-periodic
solutions was however demonstrated  for discrete systems in the
infinite domain,
and their generalization to multipulse type configurations was
also  considered.


\end{multicols}

\begin{figure}
\epsfxsize=6cm
\centerline{\epsffile{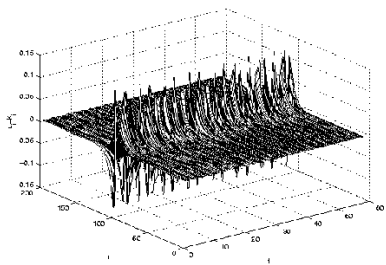}}
\caption{BOK background: Discrete SG simulation result for time
evolution of an initial condition consisting of a kink plus a projection to the
edge mode. Shown is the time evolution of the initial condition
  when substracting
$K_{gs}$ from the profile. The persistent oscillations of non-decreasing
amplitude indicate that the BOK is indeed the asymptotic state.}
\label{fig1}
\end{figure}

\begin{figure}
\epsfxsize=6cm
\centerline{\epsffile{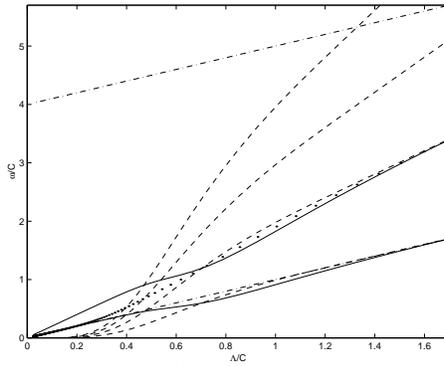}}
\caption{Internal Eigemodes, harmonics and combinations for the DNLS
case: The translational mode and its first  few harmonics are shown in dashed
line, the edge mode and its second harmonic shown in solid line, the
sum of the two frequencies is given by the dotted line. The band edges
of the PB are shown by dash-dot lines.}
\label{fig2}
\end{figure}

\begin{figure}
\epsfxsize=7cm
\centerline{\epsffile{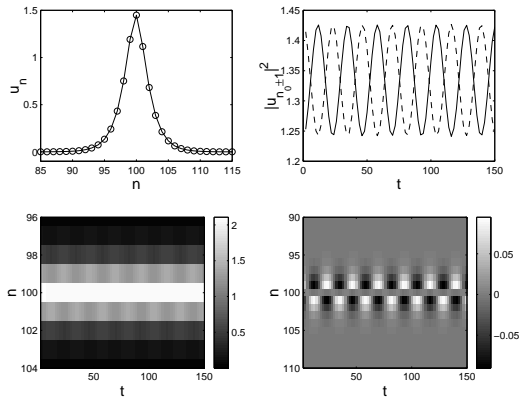}}
\caption{Example of a metastable BOB: For the DNLS and $\Lambda/C=0.370$
the time evolution of a BOB is shown. The top left panel shows the
initial condition consisting of the exact breather plus a perturbation 
along the direction of the eigenvector of the translational eigenmode.
The top right panel shows the time evolution of the amplitude squared of
the adjacent to the central
($n_0=100$ in this case) lattice sites clearly indicating the persistence
of oscillations on top of the original breather. The bottom left panel
shows the contour plot of the spatio-temporal evolution of the full BOB
while the bottom right shows the spatio-temporal evolution of the 
(amplitude squared) 
of the BOB minus the exact breather. This also clearly indicates the 
persistence of the metastable BOB (notice that very slow
decay is observed for the simulated times).}
\label{fig3}
\end{figure}

\begin{figure}
\epsfxsize=7cm
\centerline{\epsffile{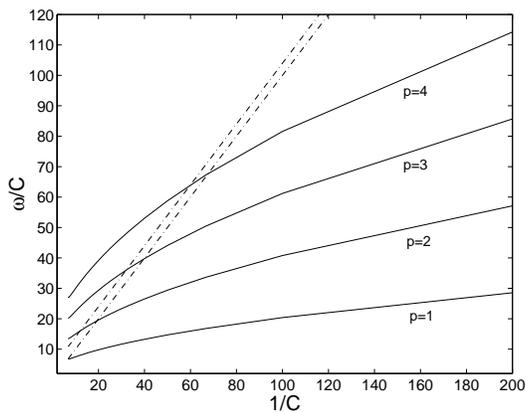}}
\caption{Relative position of the 
internal mode frequency $\omega_{\star}$ of the DKL solution 
and its $p-th$ harmonics (solid lines) with
respect to PB (dash-dotted lines) for different values of the 
coupling constant $C$. $\Lambda=1$.}
\label{fig4}
\end{figure}

\begin{figure}
\epsfxsize=7cm
\centerline{\epsffile{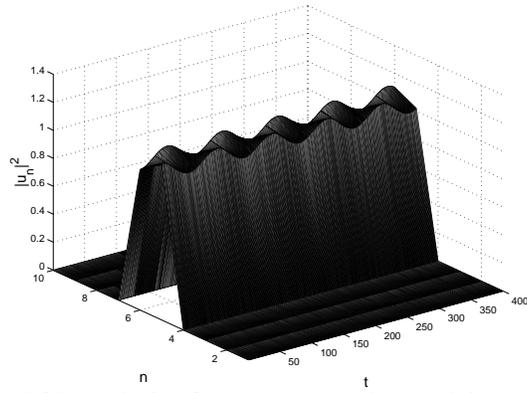}}
\caption{Proof of persistence of DKL BOB mode for $C=0.02$. The 
additional frequency of the oscillations clearly observed at the
top is checked to agree with the predictions of linear stability
theory for $\omega_{\star}$. Shown is the spatio-temporal evolution of
the amplitude squared ($|u_n(t)|^2$) of the field (see eqn. (\ref{tprleq6a})).}
\label{fig5}
\end{figure}

\newpage

\begin{figure}[tbp]
\epsfxsize=8.35cm
\epsffile{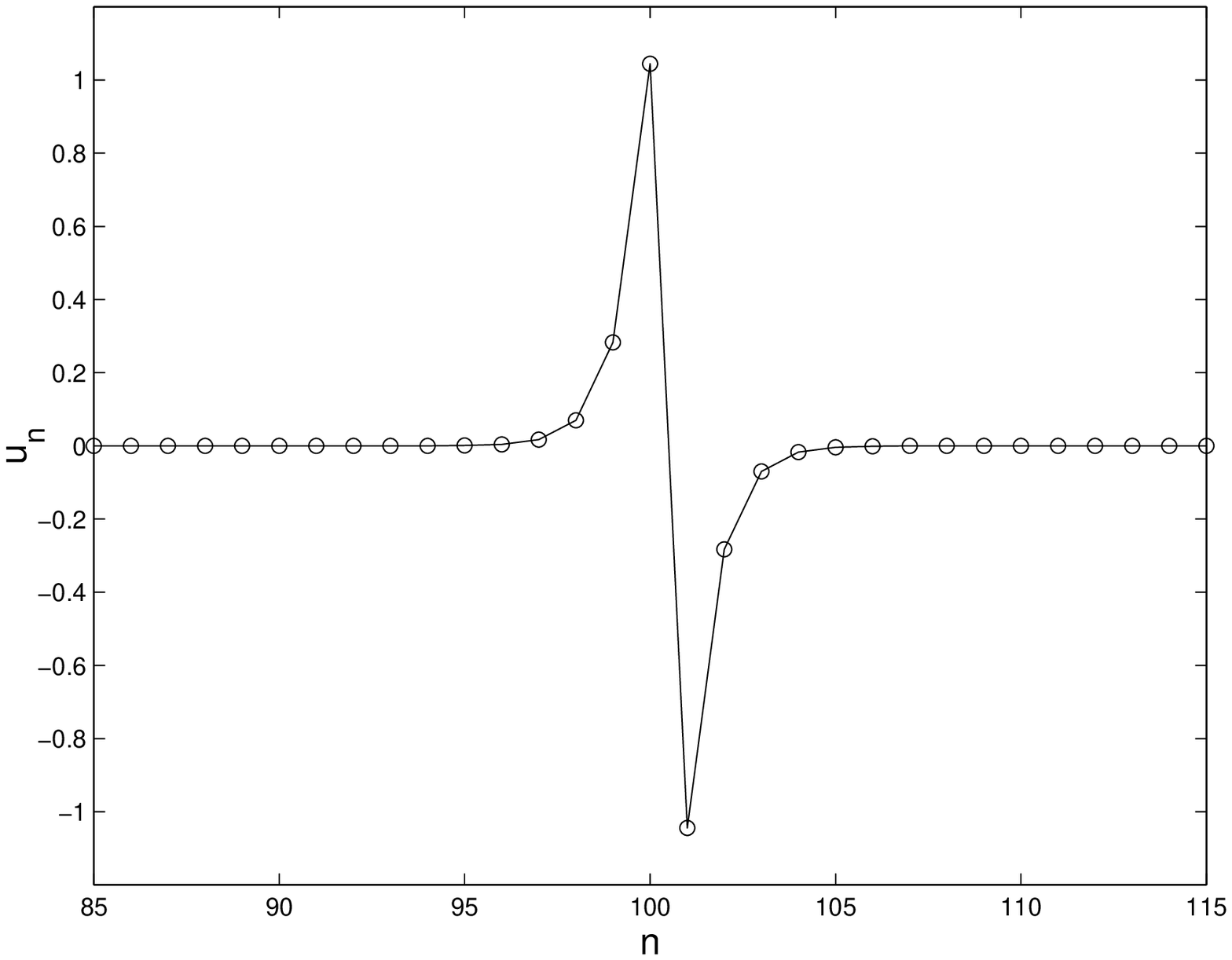}
\epsfxsize=8.35cm
\epsffile{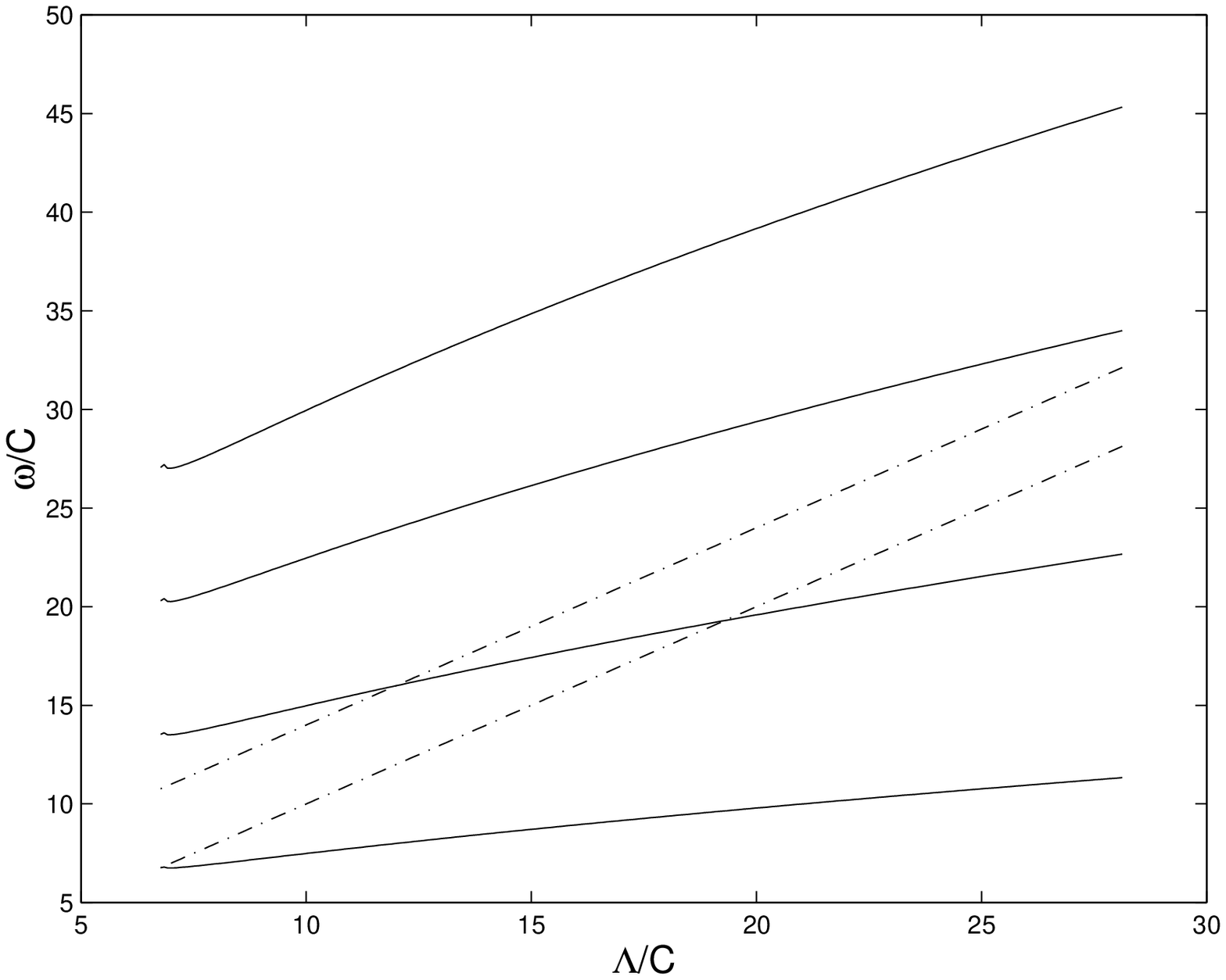}
\epsfxsize=8.35cm
\epsffile{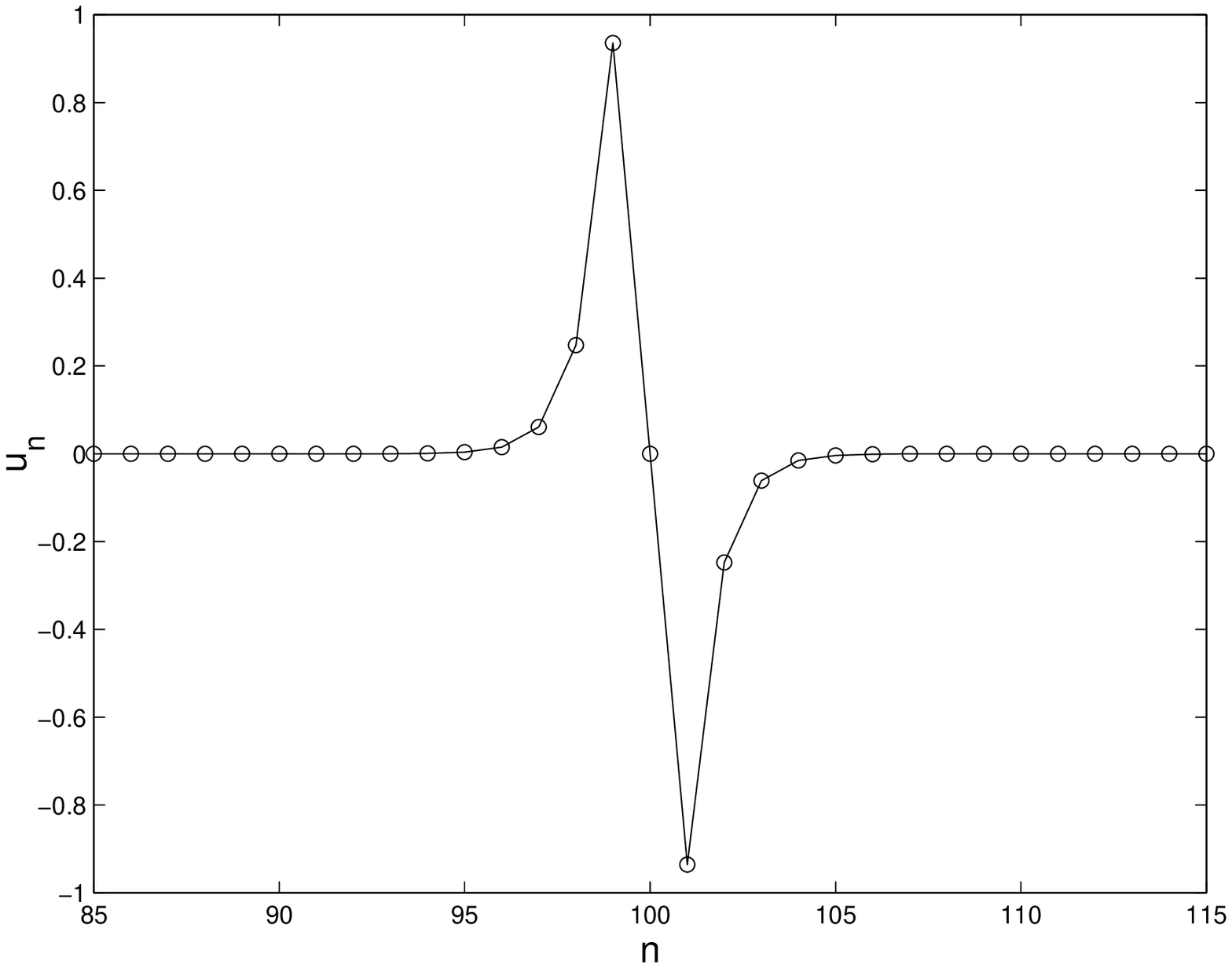}
\epsfxsize=8.35cm
\epsffile{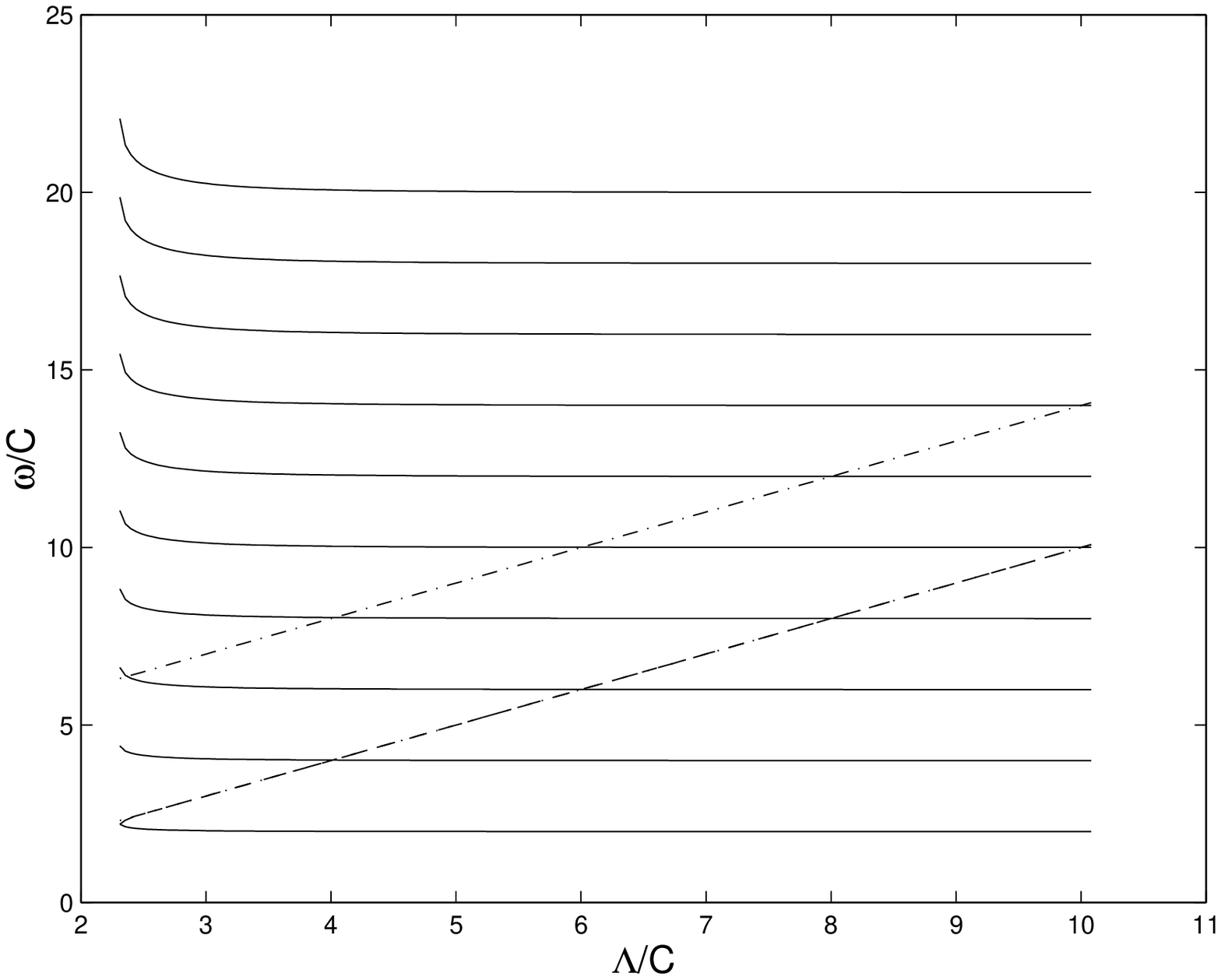}
\epsfxsize=8.35cm
\epsffile{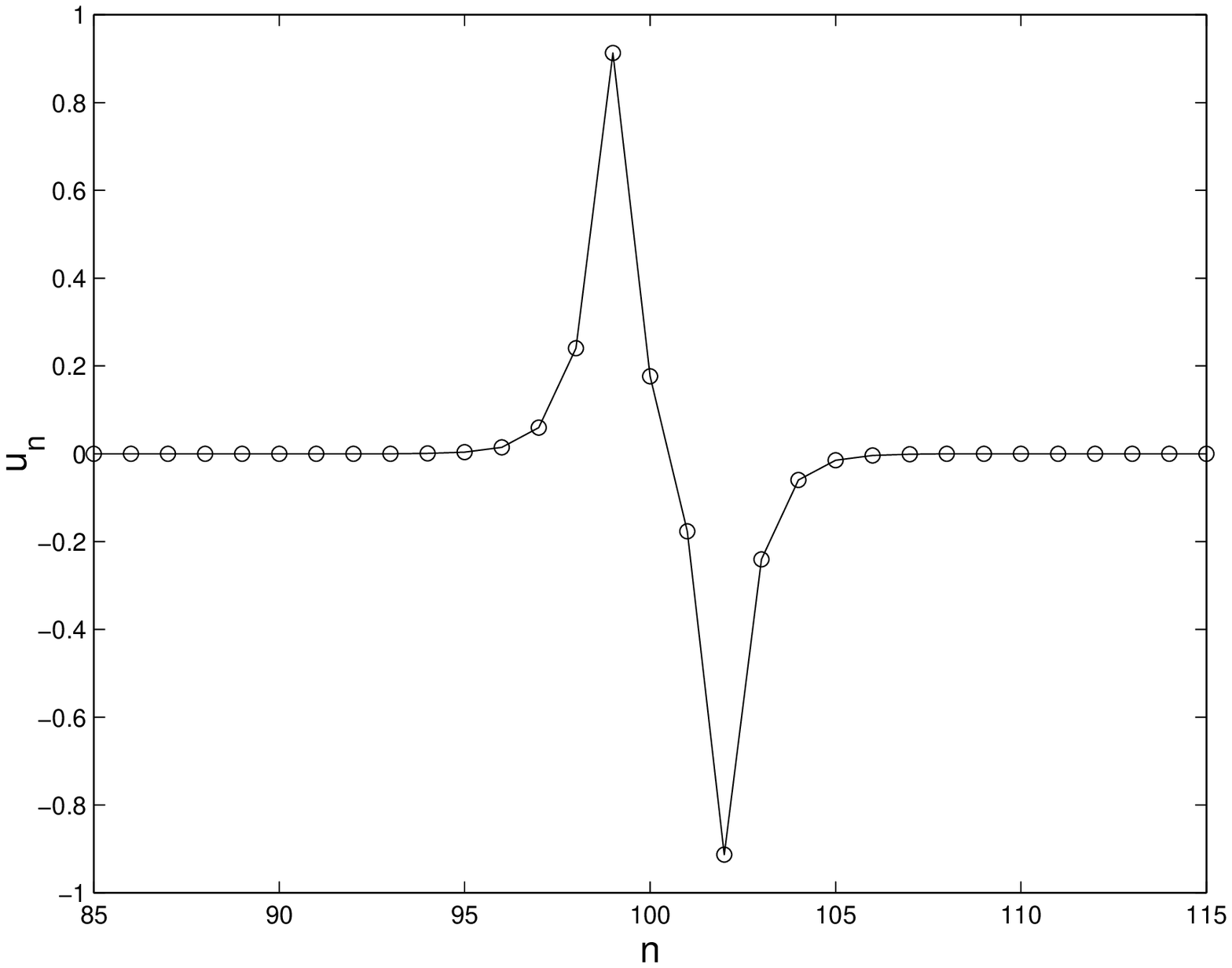}
\epsfxsize=8.35cm
\epsffile{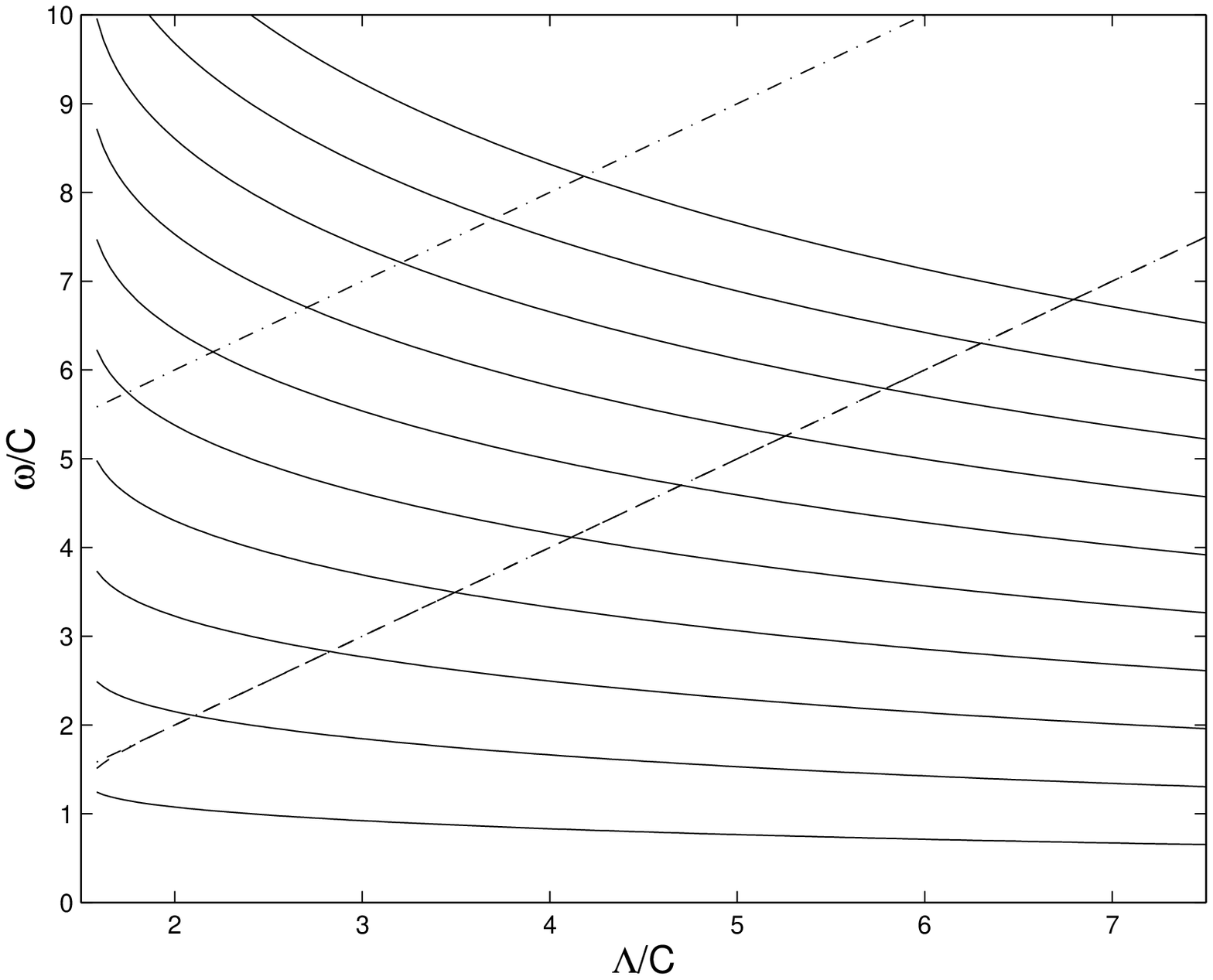}
\caption{The generalization of the twisted modes and their respective
resonance pictures. The top left panel shows the spatial profile of
the TLM and the top right, the frequency its internal mode and its first few
harmonics (given by solid line) with the PB (the edges of which are
given by dash-dotted lines. The middle row shows the extension of the
TLM for distance between centers equal to $L=2 h$ (see also text);
once
again the left panel shows the spatial profile, while the right shows
the internal frequency and its harmonics positions with respect to PB.
The same plots are shown in the bottom row for a TLM with $L=3 h$. 
It is clear that in both the middle and bottom right panels the 
decreasing dependence of the interaction mode frequency (divided by
$C$) as a function of the inverse coupling constant {\it generically}
allows for the occurrence of resonances and hence BOB type solutions
can only be metastable. $\Lambda=0.5$.}
\label{fig6}
\end{figure}

\begin{figure}[tbp]
\epsfxsize=8.35cm
\epsffile{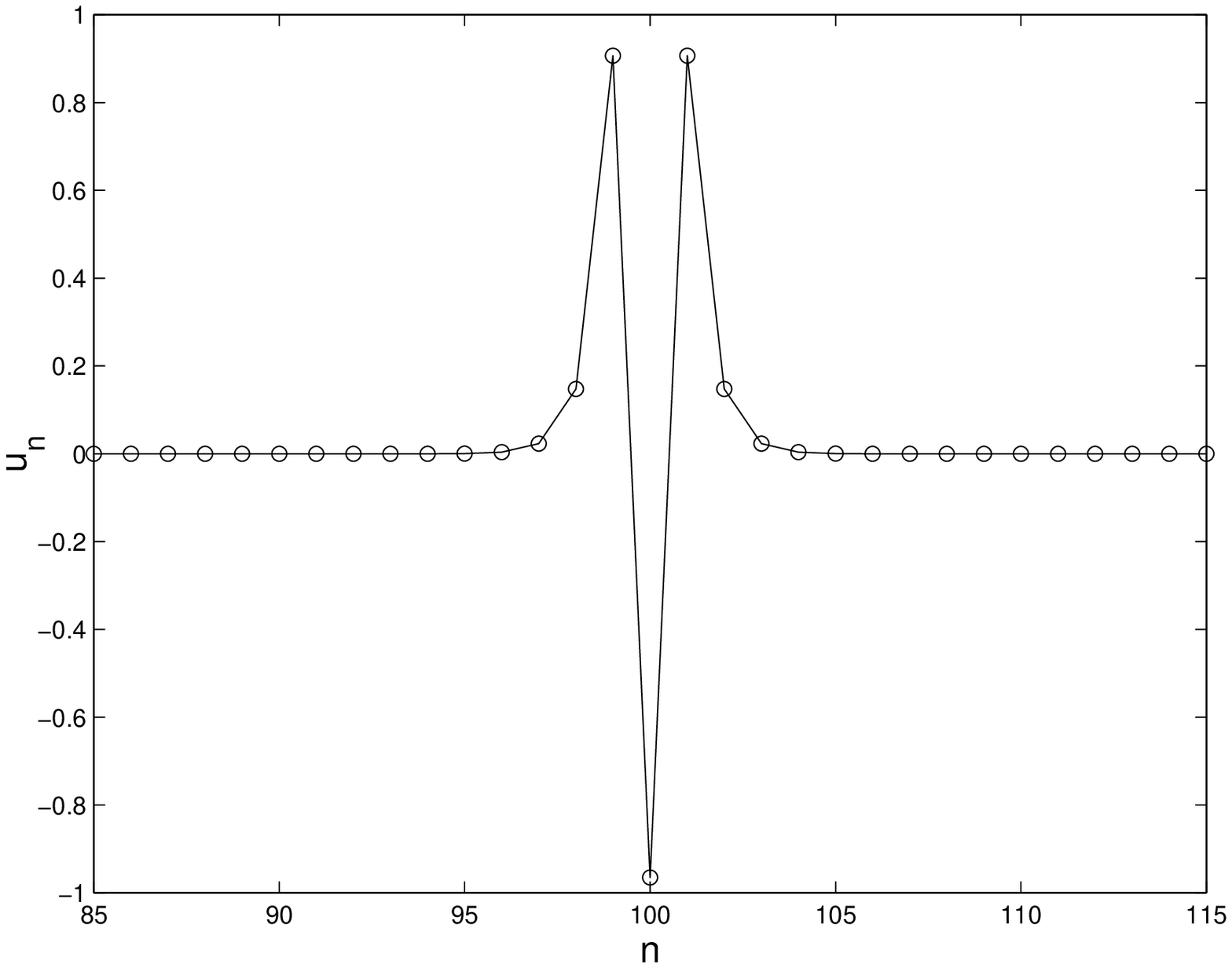}
\epsfxsize=8.35cm
\epsffile{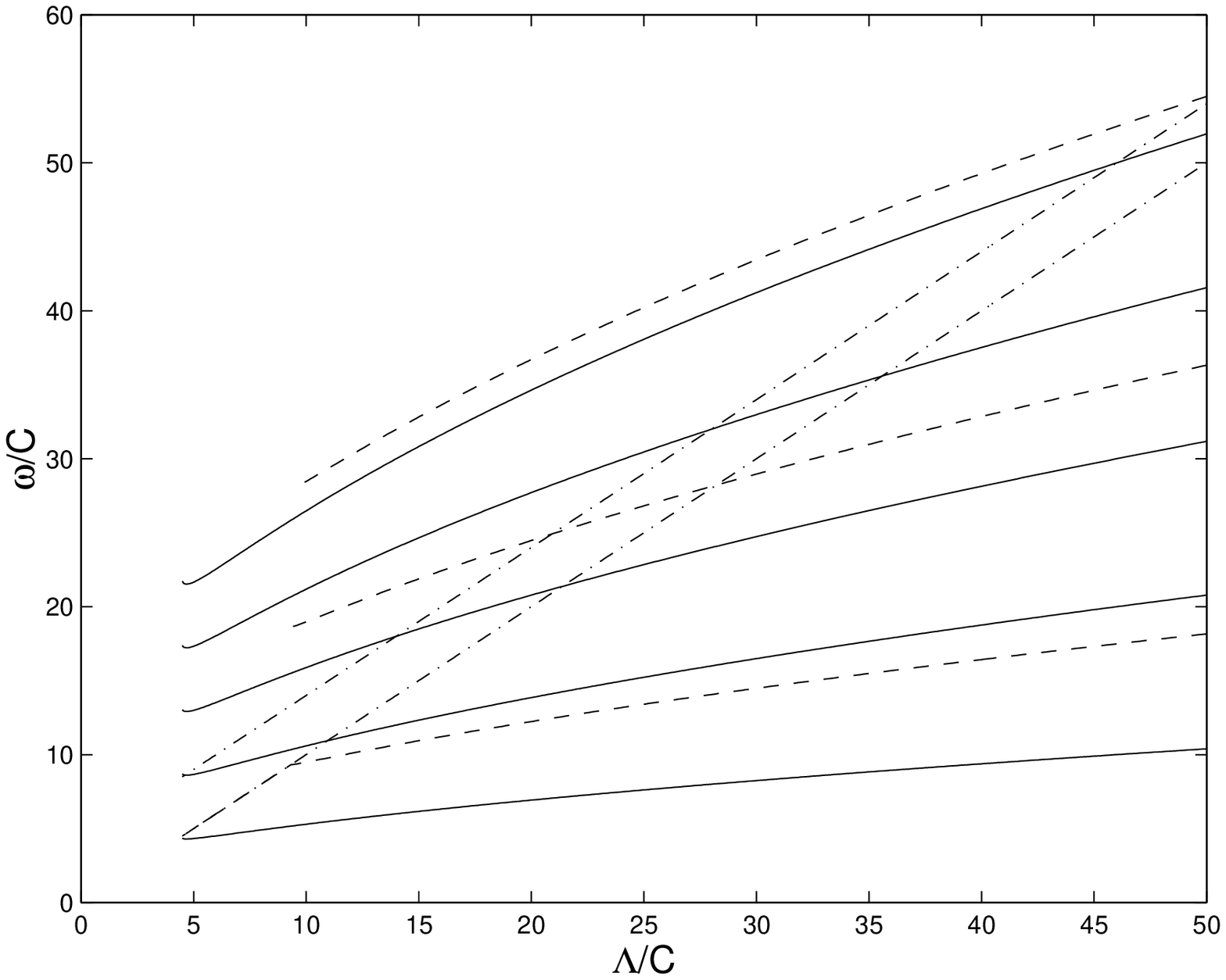}
\epsfxsize=8.35cm
\epsffile{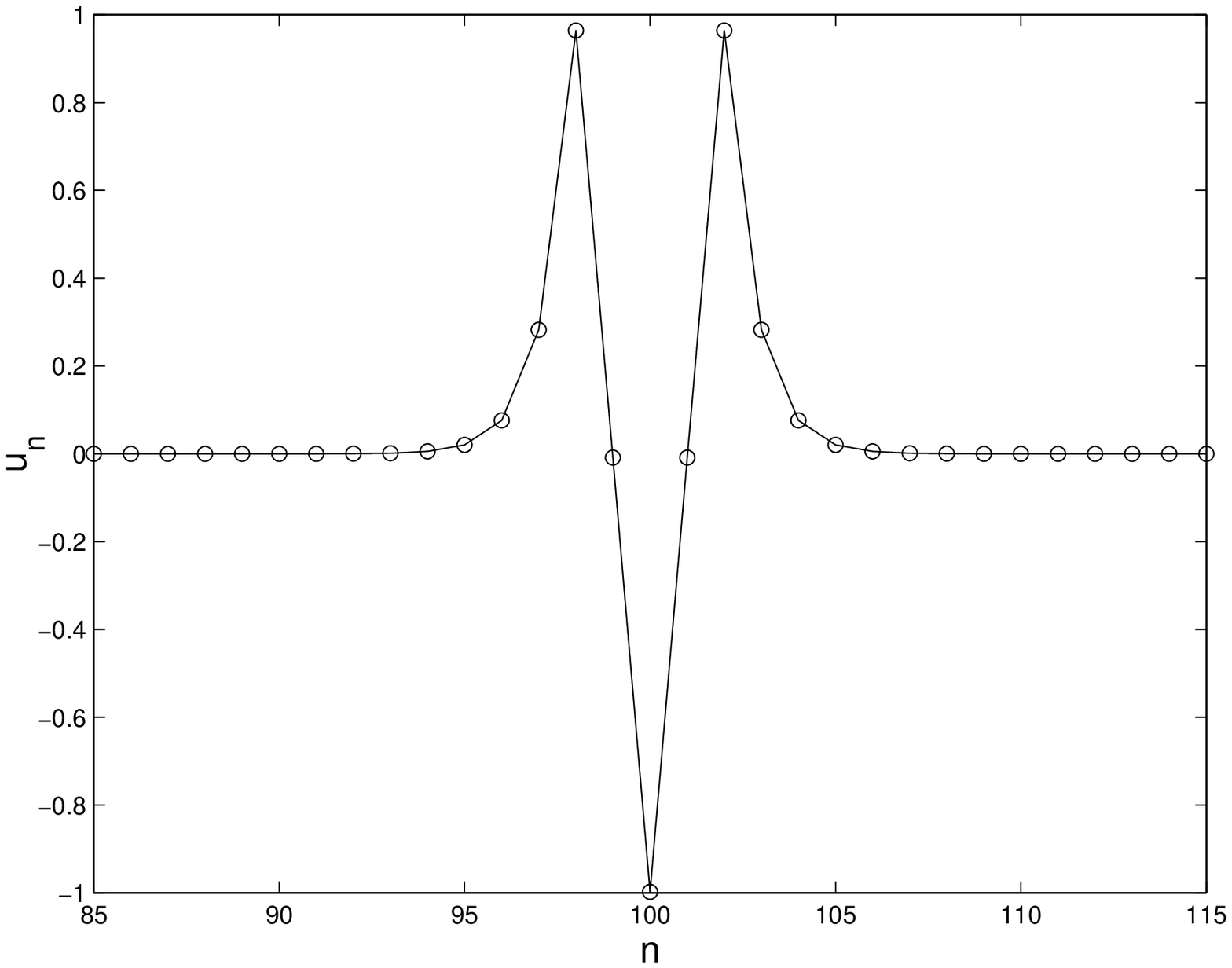}
\epsfxsize=8.35cm
\epsffile{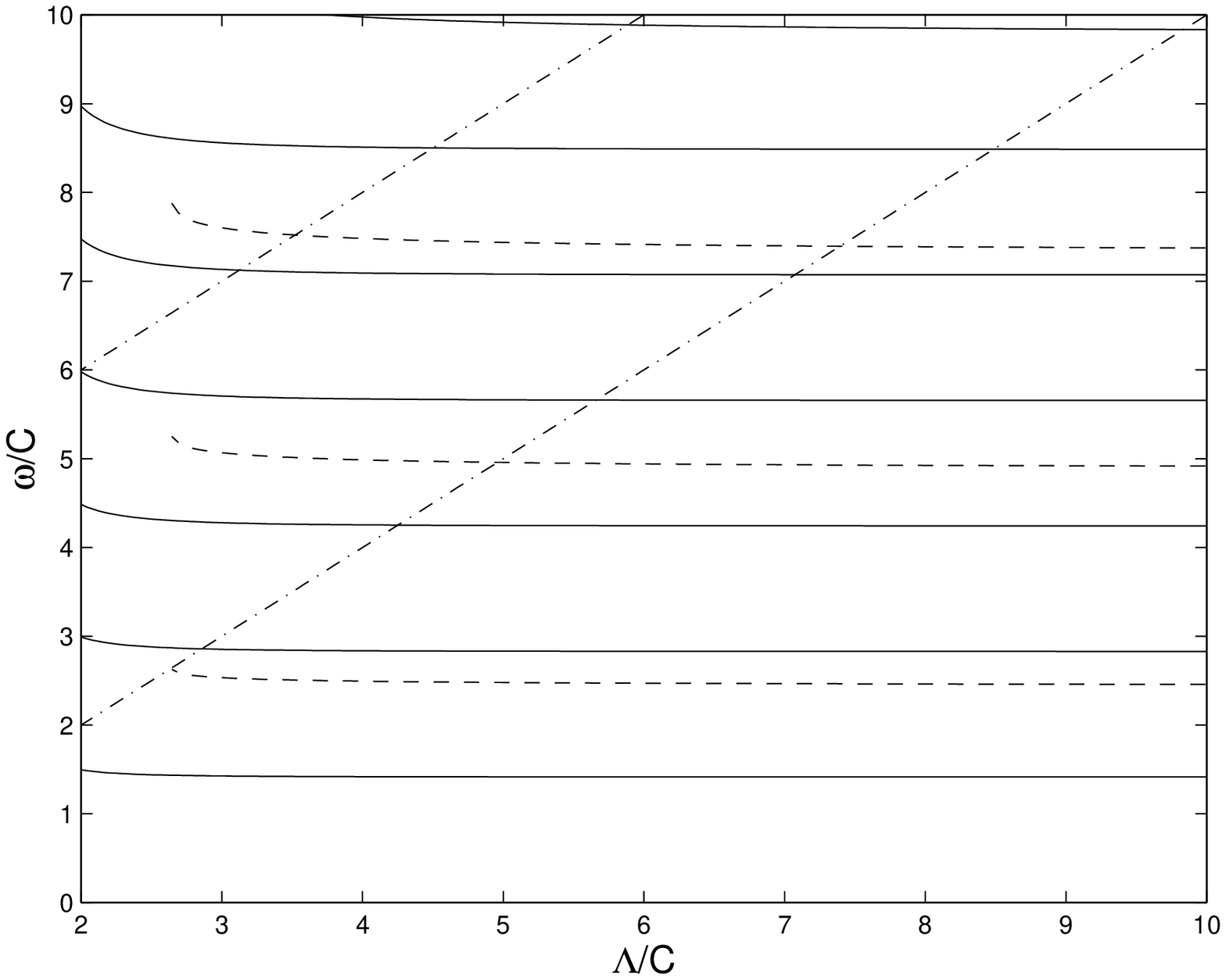}
\epsfxsize=8.35cm
\epsffile{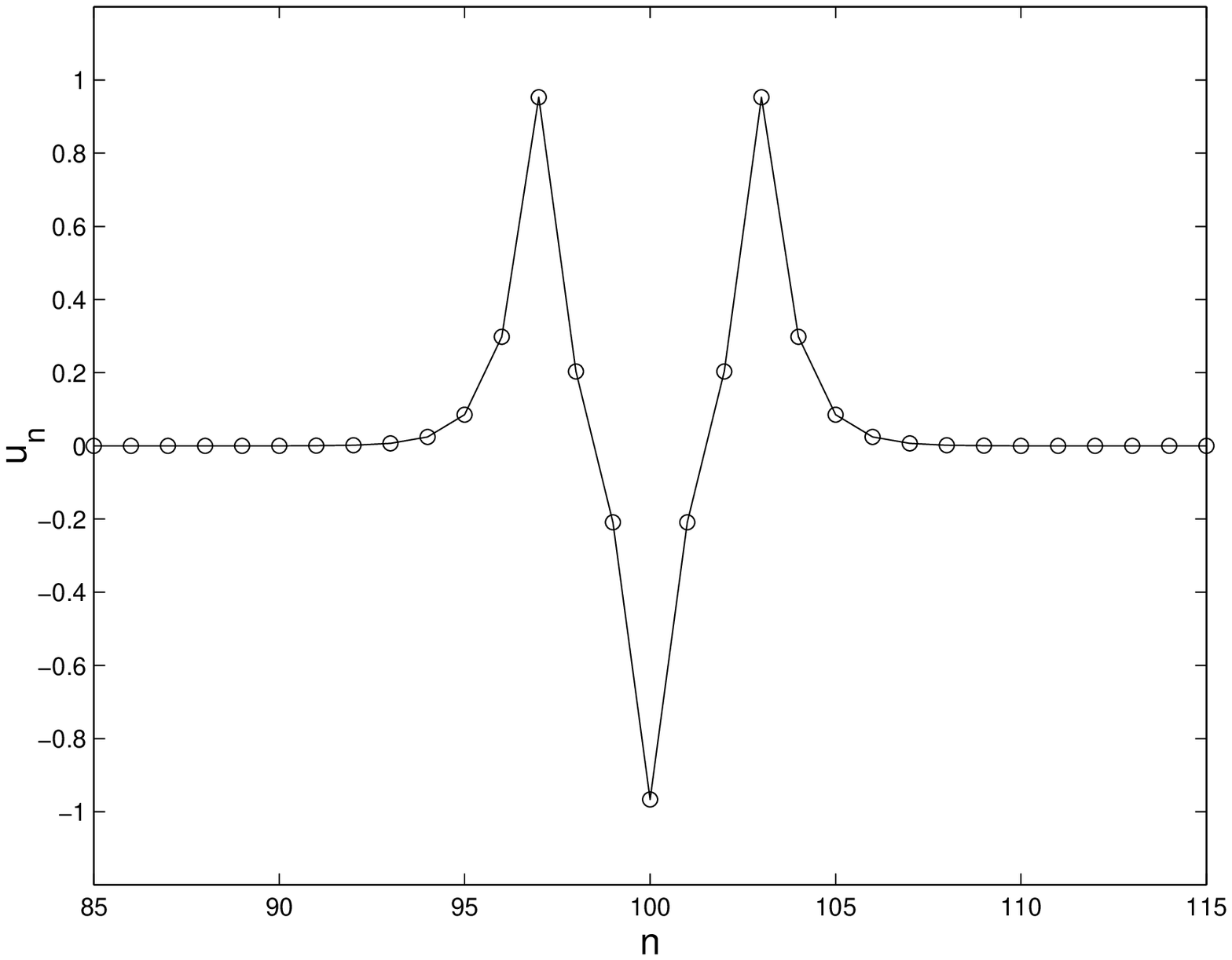}
\epsfxsize=8.35cm
\epsffile{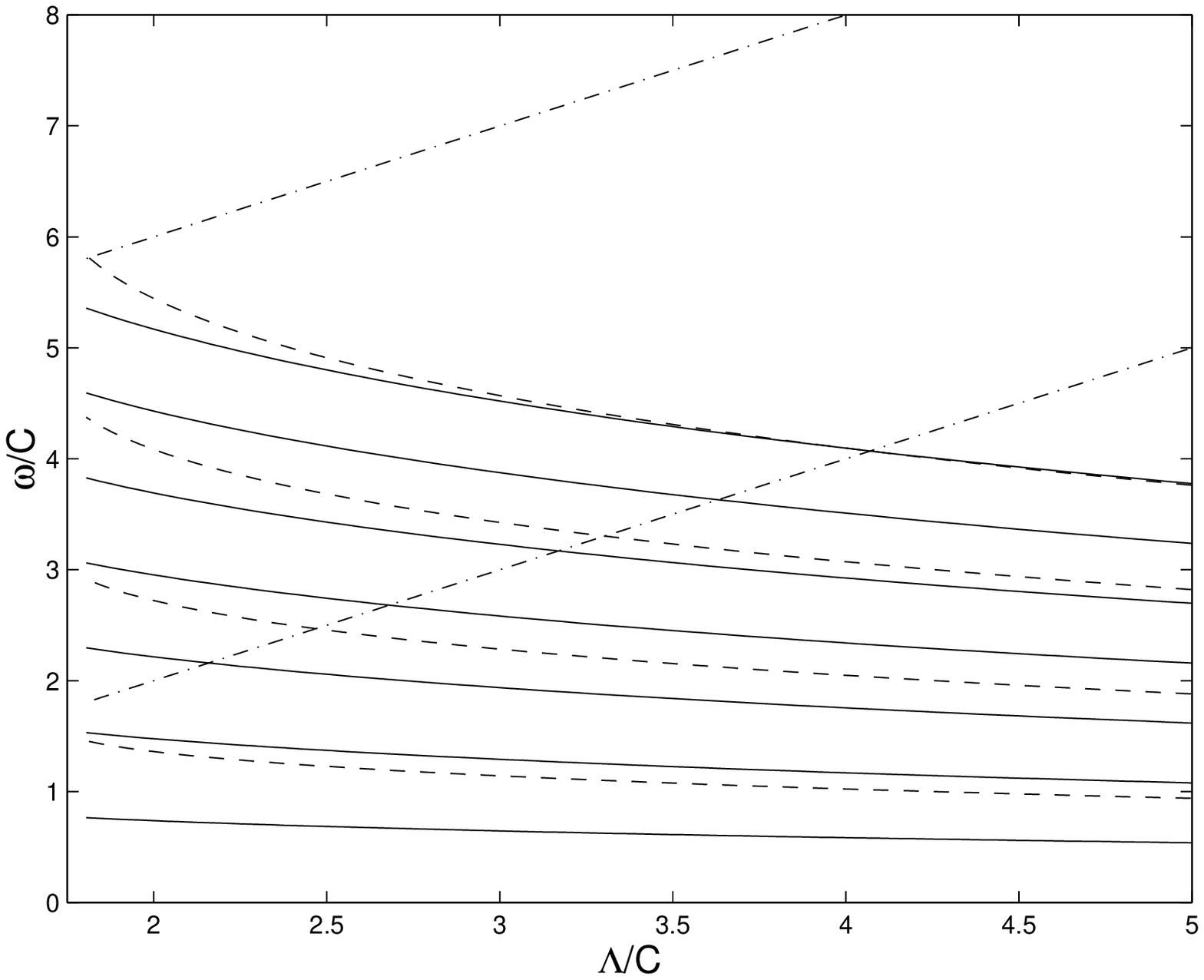}
\caption{Similar plots as the ones of Fig. 6 are given here
for the generalization to three pulses. The mode with $L=h$
is given in the top left panel while the resonance picture of
its two internal (interaction) eigenmodes (given by solid and
dashed line respectively) and their harmonics with respect to
to the PB (lying between the two dash-dotted lines) is given 
in the top right panel. Similar plots for the $L=2 h$ three
and the $L=3 h$ three pulse solutions are given in the middle
and bottom row. Notice once again that {\it only} for the
$L=h$ case there is an increasing dependence of $\omega/C$
as a function of $\Lambda/C$ for the internal modes and hence
there is a possibility for genuinely quasiperiodic solutions.
As discussed also in the text for $L \geq 2 h$, the decreasing
nature of the above dependence generates resonances with the
PB and hence becomes prohibitive for the presence of (other
than metastable) such solutions.}
\label{fig7}
\end{figure}


\newpage

{\footnotesize
\begin{table}
\caption{SG normal form, equilibria and anticipated coherent structures}
\vbox{\hfil
\begin{tabular}{|l|l|l|} \hline
Regime of $d$& Equilibria & Coherent structures\\ \hline
I:\ $d<d_e, d_e \sim 0.515$ & $P\ge0$ & $K_{gs},\ gW$\\ \hline
II:\ $d_e\le d <0.565$& $\{(P,0),(0,Q): P\ge0,\ Q\ge0\}$& $K_{gs},\ W,\ 
gW$\\ \hline
III:\ $0.565\leq d < d_*$  & $\{(0,Q):Q\ge0\}$& $K_{gs},\ W$
  \\ 
 $d_*\sim 0.86$ & &\\ \hline
IV:\ $d_* \leq d$ & $\{ (0,0)\}$ &$K_{gs}$
  \\ \hline
\end{tabular}
\hfil}
\end{table}
}

{\footnotesize
\begin{table}[h]
\caption{$\phi^4$ normal form, equilibria and anticipated
coherent structures}
\vbox{\hfil
\begin{tabular}{|l|l|l|} \hline
Regime of $d$& Equilibria & Coherent structures \\ \hline
I:\ $ d < 0.5398$  &$\{(P,0):P\ge0\}$  &$K_{gs},\ W, gW$\\
                          &$\{(0,Q):Q\ge0\}$  &
  \\ \hline
II-III:\ $0.5398\leq d < 0.6364$  &$\{(0,Q):Q\ge0\}$  &$K_{gs},\ W$ 
  \\ \hline
IV:\ $0.6364 \leq d<0.6679$ &$\{(0,Q):Q\ge0\}$  &$K_{gs},\ W,\ gW$\\
                            &$\{(P,0):P\ge0\}$ &
  \\ \hline
V:\ $0.6679 \leq d <d_e$ &  $\{(0,Q):Q\ge0\}$  & $K_{gs},\ W$\\
 $d_e\sim0.82$ & & \\ \hline
VI:\ $d_e \leq d <0.9229 $ & $\{(0,Q,0):Q\ge0\}$  &$K_{gs},\ W,\ eW$\\ 
                        & $\{(0,0,R):R\ge0\}$ &  \\ \hline
VII:\ $0.9229\leq d<1.2234$ & $\{(0,0,R):R\ge0\}$ &$K_{gs},\ eW$ \\
$d_*\sim 1.2234$ & & \\ \hline
VIII:\ $d_*\le d $ & $\{(0,0,0)\}$ &$K_{gs}$\\ \hline
\end{tabular}
\hfil}
\end{table}
}

\end{document}